\documentclass[5p,times]{elsarticle} 
\usepackage{amsmath,amssymb,bm}
\usepackage{siunitx}
\DeclareSIUnit\clight{c}
\DeclareSIUnit{\ppmV}{ppm_V} 

\usepackage{graphicx}
\usepackage{subcaption}
\usepackage{booktabs}
\usepackage{multirow}
\usepackage{microtype}
\usepackage{xspace}
\usepackage{float} 
\usepackage{balance}
\usepackage{wasysym}

\usepackage[normalem]{ulem}
\usepackage[dvipsnames]{xcolor}

\usepackage[colorlinks=true,allcolors=blue,linktocpage]{hyperref}

\biboptions{sort&compress}
\journal{Nuclear Instruments and Methods in Physics Research Section~A}

\newcommand{\pp}{\ensuremath{\mathrm{p\kern-0.05em p}}\xspace}

\newcommand{\figref}[1]{Fig.~\ref{#1}}
\newcommand{\Figref}[1]{Figure~\ref{#1}}
\newcommand{\secref}[1]{Section~\ref{#1}}

\newcommand{\tabref}[1]{Table~\ref{#1}}
\newcommand{\Tabref}[1]{Table~\ref{#1}}

\newcommand{\ArCOtwo}{Ar-CO$_2$ (90-10)\xspace}

\newcommand{\xray}{x-ray\xspace}
\newcommand{\xrays}{x-rays\xspace}

\begin{document}

\begin{frontmatter}

\title{Impact of trace amounts of water on the stability of\\ Micro-Pattern Gaseous Detectors measured in \ArCOtwo}

\author[inst1]{H.~Fribert\corref{cor1}}
\ead{henrik.fribert@tum.de}
\author[inst1]{L.~Fabbietti}
\author[inst2,inst3]{P.~Gasik}
\author[inst1]{B.~Ulukutlu}

\cortext[cor1]{Corresponding author.}

\affiliation[inst1]{organization={Technische Universität München, TUM School of Natural Sciences},
  city={Garching}, country={Germany}}

\affiliation[inst2]{organization={GSI Helmholtzzentrum für Schwerionenforschung GmbH (GSI)},
  city={Darmstadt}, country={Germany}}

\affiliation[inst3]{organization={Facility for Antiproton and Ion Research in Europe GmbH (FAIR)},
  city={Darmstadt}, country={Germany}}

\begin{abstract}
In this study, we investigate the influence of humidity on the performance of various non-resistive Micro Pattern Gaseous Detectors, such as GEM, Thick-GEM, and Micromegas, operated with \ArCOtwo gas mixture. The water content is introduced in a range of \SIrange[range-units=single]{0}{5000}{\ppmV}. It is observed that the presence of increased humidity does not significantly degrade any of the studied performance criteria. On the contrary, our measurements suggest an improvement in discharge stability with increasing humidity levels at the highest gains and fields. No significant difference is observed at the lower gains, indicating that humidity helps to reduce the rate of spurious discharges related to electrode defects or charging-up of the insulating layers. We conclude that adding a small amount of water to the gas mixture may be beneficial for the stable operation of an MPGD.
\end{abstract}

\begin{keyword}
MPGD \sep GEM \sep THGEM \sep Micromegas \sep discharge \sep humidity
\end{keyword}

\end{frontmatter}



\section{Introduction} 

\label{sec:intro}

\begin{figure*}[ht]
    \centering
    \includegraphics[width=0.75\linewidth]{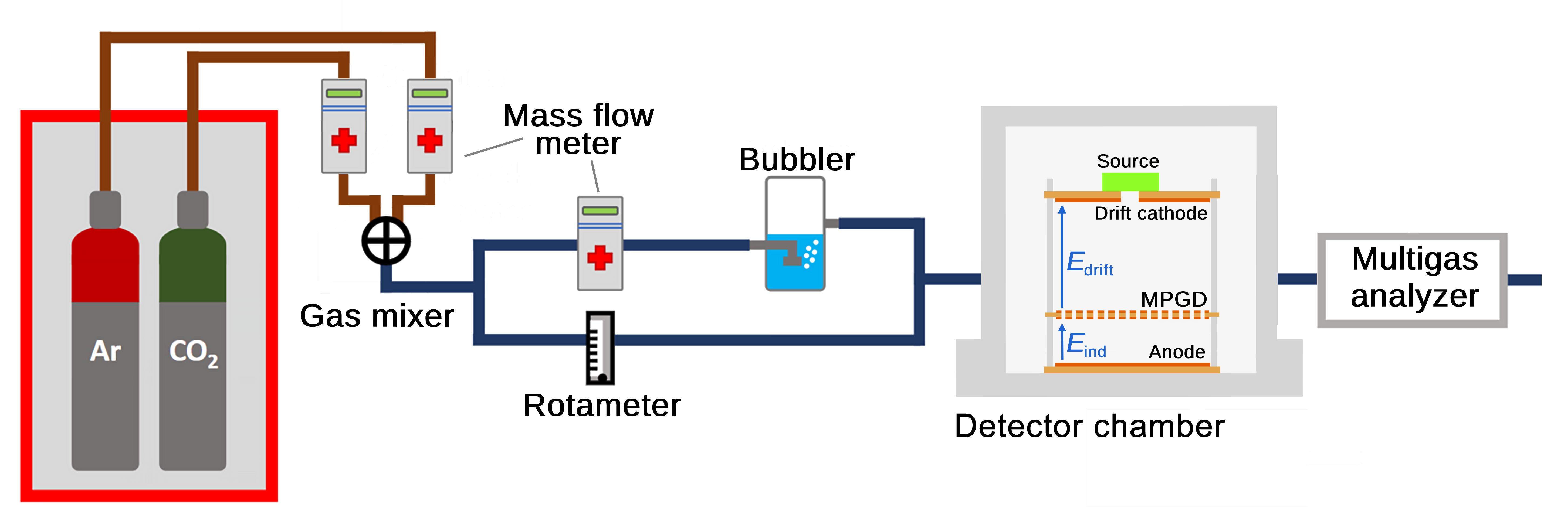}
    \caption{The experimental setup of the detector chamber and the gas system used to humidify the gas mixture.}
    \label{fig:setup}
\end{figure*}
Despite several decades of experience in the production and successful operation of Micro Pattern Gaseous Detectors (MPGDs, see recent review in \cite{sauli2020book}), the effect of water contamination in the gas composition on their performance is still a subject of debate. Previous studies regarding the addition of water vapour to the gas are inconclusive, and no consensus has been reached, especially concerning discharge stability and the influence on streamer development~\cite{ALTUNBAS2002177, SF6, en15030817}. Adding small amounts of water vapour to the gas may benefit detector operation by preventing the glue used in detector components from drying out and helping to mitigate ageing effects~\cite{HOHLMANN2002179}. However, an adverse effect of small amounts of humidity in the gas mixture on the detector performance has to be ruled out before its utilisation in experiments. 

To resolve this question, we study various MPGDs, including Gas Electron Multiplier (GEM)~\cite{SAULI1997531}, Thick Gas Electron Multiplier (THGEM\footnote{The results have been partially published in~\cite{Fribert_2023}})~\cite{CHECHIK2004303, CHECHIK200535}, and the Micro-mesh gaseous structure (Micromegas, MMG)~\cite{GIOMATARIS199629}. We investigate different performance criteria of the detectors, including gain and energy resolution, and put a particular focus on discharge stability. We examine the impact of water vapour in the gas mixture on detector performance by building a dedicated setup to introduce and vary the humidity between measurements. 

Our measurements are performed in \ArCOtwo at the absolute humidity levels of \SIrange[range-units=single]{0}{5000}{ppm_V} (parts per million, proportion by volume), which correspond to a relative humidity (RH) value of up to $\sim$20\% at normal temperature and pressure conditions.
It should be noted that several groups reported various instabilities in MPGD detectors appearing at much larger relative humidity values while operating/testing a detector structure in an air atmosphere in the laboratory. In~\cite{Ball_2017}, instabilities such as increased leakage current and spark rate were observed in GEMs operated in ambient, cleanroom conditions (RH$\approx$50\%). Similar observations have also been made with other types of MPGDs (see, for example, \cite{Iengo_MMG} for resistive Micromegas or recent results with \textmu RGroove presented in~\cite{urgroove_drd1}). For this reason, the humidity limits for the quality control procedures of MPGDs are usually put below the RH value of 50\% or even below 10\%, as required in the recently concluded large-scale productions for the LHC detector upgrades~\cite{Adolfsson_2021, ABBAS2022166716, ALLARD2022166143}. This is well in line with the range considered in this work. In the presented studies, we focus on the fundamental stability limits of MPGD-based detectors in a gas mixture with trace amounts of water, and the regions of large RH ($>50\%$) are not considered here.

\section{Experimental setup and methods}
\begin{figure}[t]\centering
    \includegraphics[width=\linewidth]{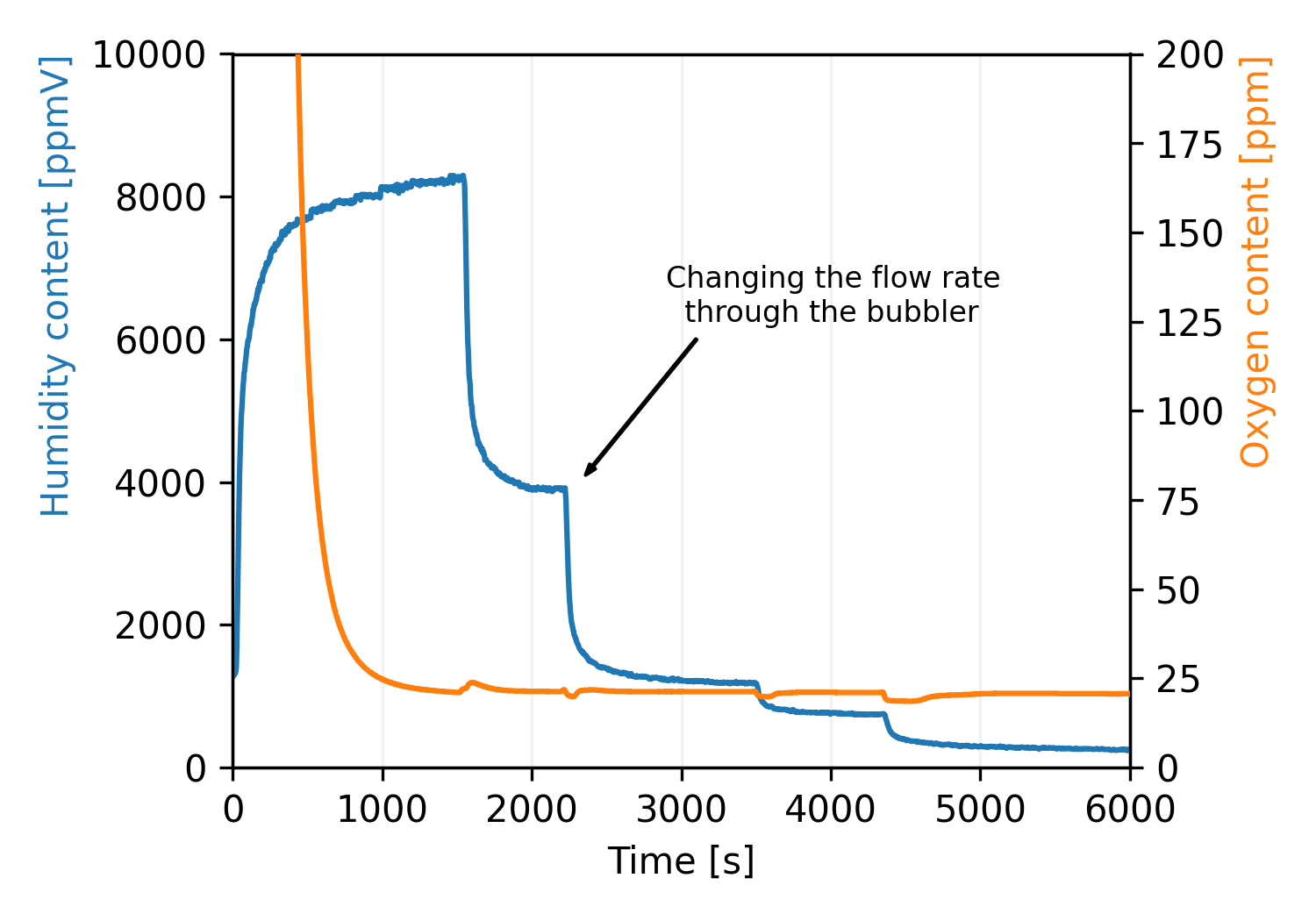}
  \caption{Gas contaminants recorded during a test measurement. The initial spike in the oxygen content corresponds to the activation of the humidifier at time zero. It is demonstrated that humidity can be controlled reliably by changing the flow rate through the bubbler.}
  \label{fig:humidity}
\end{figure}
\subsection{Gas mixture}
\vskip5pt
The gas mixture used in the presented studies is \ArCOtwo.
A mixture of argon and carbon dioxide is widely employed in various ratios in modern MPGD-based detectors, including the COMPASS GEM tracker~\cite{ALTUNBAS2002177}, the CMS Muon Endcap GEM~\cite{Colaleo:2021453}, and the ATLAS New Small Wheel Micromegas~\cite{Kawamoto:1552862}, among others. While argon can be considered the most common noble gas in such detectors, the use of CO$_2$ as a quencher --- though not always optimal for detector performance --- remains highly attractive due to its environmental compatibility (non-greenhouse gas), non-toxicity, non-flammability, and chemical inertness, making it particularly suitable for large-scale MPGD applications. The noble gas to quencher ratio of 90-10 used in this study was primarily chosen to allow direct comparison with a broad range of our previous measurements on MPGD stability (e.g.~\cite{GASIK2017116, Gasik:2022vft, Gasik_2023}). 

It should be noted that the effects of humidity observed in this study may not be directly transferable to other gas mixtures, as discharge properties and quenching mechanisms can vary significantly. Furthermore, in mixtures containing fluorinated gases, the presence of humidity should be strictly avoided, as it may lead to the formation of hydrofluoric acid, posing a risk to the integrity of detector components~\cite{HOHLMANN2002179, CAPEANS200373, CORBETTA2021165373}. Each gas mixture should therefore be studied individually to assess its specific behaviour under varying humidity conditions.

\subsection{Gas system}
\vskip5pt

A dedicated setup is built to study the effect of humidity on MPGDs, in which a constant amount of water vapour is reliably introduced to the gas. \Figref{fig:setup} shows a scheme of the humidification process. The original gas mixture flows through a piping system, splitting into two separate lines. One line incorporates a bubbler filled with distilled water, allowing the gas to absorb humidity without introducing additional contaminants like oxygen. Different humidity levels can be achieved by adjusting the flow through the bubbler using a mass flow meter, as shown in \figref{fig:humidity}. Simultaneously, the second line directs dry gas around the humidifying line until they merge. The gas mixture is continuously flushed through the $\sim$4-litre detector vessel at a flow rate of 10\,l/h, which is regulated by mass flow controllers and monitored using a rotameter installed in the dry gas line. A Rapidox 3100 Multigas Analyser from Cambridge Sensotec is placed after the detector vessel to record the oxygen and humidity content of the mixture. The manufacturer does not specify the precision of the humidity measurements. However, based on the reproducibility of the values obtained in this work, the uncertainty in the water content is considered negligible. The oxygen content is kept at a minimum ($<25$\,ppm) throughout all measurements. The ambient conditions are tracked with thermometers and a pressure sensor.


\begin{table*}[t]
  \centering\footnotesize
    \caption{A list of MPGD structures used in this study. The naming convention for Micromegas follows the one established in~\cite{Gasik_2023}. The main geometrical parameters are specified: $d_{\mathrm{ins}}$ - Apical\textsuperscript{\textregistered}/FR4 thickness (for GEM/THGEM, respectively), $d_{\mathrm{Cu}}$ - copper cladding thickness, \diameter - hole diameter (inner/outer), the hole pitch, $z_{\mathrm{MMG}}$ - amplification gap given by the thickness of the Pyralux\textsuperscript{\textregistered} PC1025 pillars, $d_{\mathrm{wire}}$ - wire thickness, $a_{\mathrm{wire}}$ - distance between the wire edges (inner dimensions of a mesh cell), density in lines per inch, optical transparency. GEM (S) and GEM (LP) denote two variants, with (S) referring to the standard geometry and (LP) to the large-pitch version. All meshes are woven. The THGEM has no rims around the holes.}
    \begin{tabular}{lcccccccccc}
      \toprule
      \multirow{3}{*}{MPGD} & \multicolumn{4}{c}{(TH)GEM} & \multicolumn{4}{c}{Micromegas}  &  & \\ \cmidrule(l){2-5}\cmidrule(l){6-9}
      & $d_{\mathrm{ins}}$ & $d_{\mathrm{Cu}}$ & $\diameter$ & Pitch & $z_{\mathrm{MMG}}$ & $d_{\mathrm{wire}}$ & $a_{\mathrm{wire}}$ & Density & \shortstack{Optical\\transparency}
 & Producer \\[.5ex]
      & [\textmu m] & [\textmu m] & [\textmu m] & [\textmu m] & [\textmu m] & [\textmu m] & [\textmu m] & [LPI] & [\%] & \vspace{0.5ex}\\\midrule\\[-1.5ex]
      GEM (S) & 50  & 5 & 50/70 & 140 & -- & -- & -- & -- & 24 & CERN \\
      GEM (LP) & 50  & 5 & 50/70 & 280 & -- & -- & -- & -- & 6 & CERN \\
      THGEM  & 400  & 35 & 400 & 800 & -- & -- & -- & -- & 24 & Eltos S.p.A. \\
      MMG1 & --  & -- & -- & -- & 128 & 13 & 22 & 730 & 40  & CERN \\
      MMG3 & --  & -- & -- & -- & 128 & 18 & 45 & 400 & 51 & CERN \\
      \bottomrule
    \end{tabular}
  \label{tab:mpgd}
\end{table*}

\subsection{MPGDs}
\vskip5pt

The detector setup follows the same scheme for all measurements. It consists of an MPGD mounted between a drift cathode and a readout plane. \Tabref{tab:mpgd} includes a list of all MPGD structures used in this study, together with the main geometrical parameters.
GEM and Micromegas structures are $10\times10$\,cm$^2$ large, whereas the area of the THGEM is $11.2\times11.2$\,cm$^2$. All structures are non-resistive.

The drift cathode and the readout anode are both made of \SI{1.5}{\mm} thick PCB, coated on one side with copper. If not stated otherwise, the distance between the source and the MPGD structure is \SI{32}{\mm}, and in the case of the (TH)GEM, the induction gap length is \SI{2}{\mm} throughout all measurements. The distance $z_{\mathrm{MMG}}$ between a Micromegas mesh and the readout anode is 128~\textmu m, as indicated in \tabref{tab:mpgd}.

\subsection{Radioactive sources}
\vskip5pt

\label{sec:src}
Two different sources are used to measure and compare the performance. The first one is a mixed alpha source containing $^{239}$Pu, $^{241}$Am and $^{244}$Cm. The emitted particles have weighted mean energies of $\SI{5.15}{\mega\electronvolt}$, $\SI{5.45}{\mega\electronvolt}$ and $\SI{5.80}{\mega\electronvolt}$, respectively~\cite{alpha}. The source rate measured with the setup is $\sim$ \SI{300}{\Hz}. The second source is $^{55}$Fe, which emits mainly \xrays with an energy of $\SI{5.9}{\kilo\electronvolt}$ and a rate of $\sim$\SI{120}{\kilo\Hz}. For the discharge stability studies, the source is placed on top of the drift cathode, irradiating the drift volume through a 7.8~mm hole perpendicular to the MPGD structure.

\subsection{Gain measurements}\vskip5pt
\label{sec:gain}
The absolute gain is defined as the ratio of the amplification current to the primary current measured in the drift volume. The currents are measured using a floating multi-channel picoamperemeter~\cite{UTROBICIC201521}. For the (TH)GEM, this process is carried out with zero induction field, while the amplified signal from avalanche multiplication is read at the bottom of the foil. The primary current is measured at the top (TH)GEM electrode, with the bottom electrode and the anode plane grounded. A drift field of \( E_\mathrm{drift} = \SI{400}{\volt\per\centi\meter} \) is applied between the cathode and the GEM foil, at which the electron collection efficiency of primary charges is expected to be nearly 100\%~\cite{GASIK2017116, Gasik:2022vft, BACHMANN1999376, ratzaphd}. The measurement with a Micromegas is performed similarly by applying a drift field of $E_\mathrm{drift}=\SI{400}{\volt\per\centi\meter}$ and keeping the mesh and the anode plane grounded. Since the electron transparency of MMG1 is known to be less than 100\% for the measured field configurations, the gain values are corrected using data obtained with this specific Micromegas structure~\cite{Gasik_2023}, where electron transparency is measured as a function of the ratio \( E_\mathrm{drift}/E_\mathrm{MMG} \), with \( E_\mathrm{MMG} \) representing the field strength in the amplification region. For MMG3, a 100\% collection efficiency can be safely assumed~\cite{Gasik_2023}.

During all measurement sessions, the primary current is measured \SI{15}{\minute} after the drift voltage is applied. Each measurement lasts \SI{5}{\minute}, during which the primary current values are averaged. Subsequently, the voltage --- and consequently, the gain --- is gradually increased, waiting approximately 5 minutes after applying the voltage before measuring the amplification current for \SIrange{1}{2}{minutes} at each step, continuing until reaching a region where discharges may occur.
This means that between applying a potential to MPGDs and measuring first discharges, approximately 45 minutes pass, which is comparable with the typical charging-up time constants measured in various setups~\cite{Adolfsson_2021, Gasik:2022vft, HAUER2020164205, Pitt_2018, ALEXOPOULOS2011110, Alexeev_2015} and should be adequate to assure an MPGD structure to be sufficiently charged-up. However, it should be noted that no detailed studies on the influence of humidity on the charging-up behaviour of MPGDs have been performed in the scope of this work.

\begin{figure}[h]
    \centering
    \includegraphics[width=\linewidth]{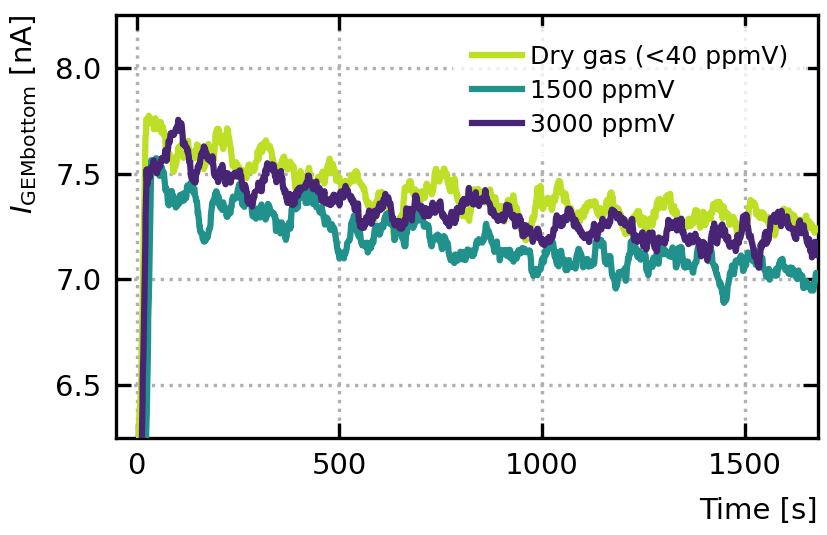}
    \caption{Current measured at the bottom side of the GEM (S) foil as a function of amplification voltage, increased from 360~V to 380~V (corresponding to a gain increase from approximately 300 to 500), shown for three different humidity levels.}
    \label{fig:currents}
\end{figure}

Upon reaching the first high-gain point, at which the first discharge signals are expected, the time between the voltage application and the start of gain and discharge probability measurement is increased to approximately 15 minutes to ensure stable gain behaviour. \Figref{fig:currents} shows an example of a GEM  amplification current measured after reaching the high-gain region, in which the first discharges occurred. 900 seconds after ramping the voltages up, the actual gain and discharge measurement starts. The amplification current is measured for the remaining time, and the average value is calculated to estimate the gain. The standard deviation of this measurement reflects the residual change of gain with time. Following the procedure described in~\cite{Gasik:2022vft}, the current excursions measured during a discharge event (current spikes) are excluded from calculating the average. 

The similar current values observed under both dry and humid (3000~ppm$\mathrm{V}$) conditions, along with the slightly lower current measured at 1500~ppm$\mathrm{V}$ water content, suggest that the observed variations are not due to a systematically increased humidity level but rather to day-to-day fluctuations in temperature and pressure—especially considering that the differences between the curves remain below 3$\%$. All three curves also exhibit similar time constants of approximately 200 seconds.
Based on these observations, we conclude that the procedure for voltage application, including the 900-second interval prior to the start of the measurement, can be applied independently of the water content in the gas.

\subsection{Amplitude spectra}\vskip5pt
\label{sec:adc}
Studies on the effect of humidity on amplitude spectra are performed to test if the detector performance may be seriously altered by the water content in the gas mixture. The measurements presented here are performed with a GEM foil, setting the induction field to \SI{4}{\kilo\volt\per\cm} and the amplification voltage to \SI{410}{\volt} to achieve a high effective gain without frequent discharging. The signal produced by the $^{55}$Fe \xray source is measured with a single, 10$\times$10\,mm$^2$ pad on the multi-pad readout anode and analysed with a simple readout chain including a pre-amplifier, an amplifier and a multi-channel analyser (MCA) for amplitude measurements and histogramming.  

Before the measurement, all components (GEM, amplifiers, MCA) are turned on and running overnight to ensure that there are no effects by charge- or warm-up of the instruments that could influence the measurements. The measurement procedure follows: recording the first spectrum for \SI{15}{\minute} in dry gas, increasing the humidity and subsequently recording the next spectra (for \SI{15}{\minute} each) until reaching $\approx{3000}$\,ppm$_{\mathrm{V}}$. At each step, it is required that the water content value stabilises (see \figref{fig:humidity}) before starting the measurement. However, the humidity level is constantly monitored, and in case it remains changing during the measurement, these residual variations are reflected in the uncertainty of the measured value. The series is then repeated, starting with humidified gas and performing a scan with decreasing water content. 

The energy resolution values are calculated by fitting a Gaussian to the main peak of the spectrum and taking the ratio of the Gaussian width ($\sigma$) to the mean peak position $\mu$. The systematic uncertainty is estimated by varying the fit ranges. The mean peak position is then corrected for the temperature and pressure ($T/p$) variations, assuming linear dependency of gain (thus $\mu$ value) on $T/p$ ratio, which was monitored throughout the measurement. It should be noted that no detailed $T/p$ calibration is available for the detectors under study. Thus, although the $T/p$ variations were rather limited during the measurement time, the residual gain dependency remains uncorrected.

\begin{figure*}[hbt]
    \centering
    \includegraphics[width=0.49\linewidth]{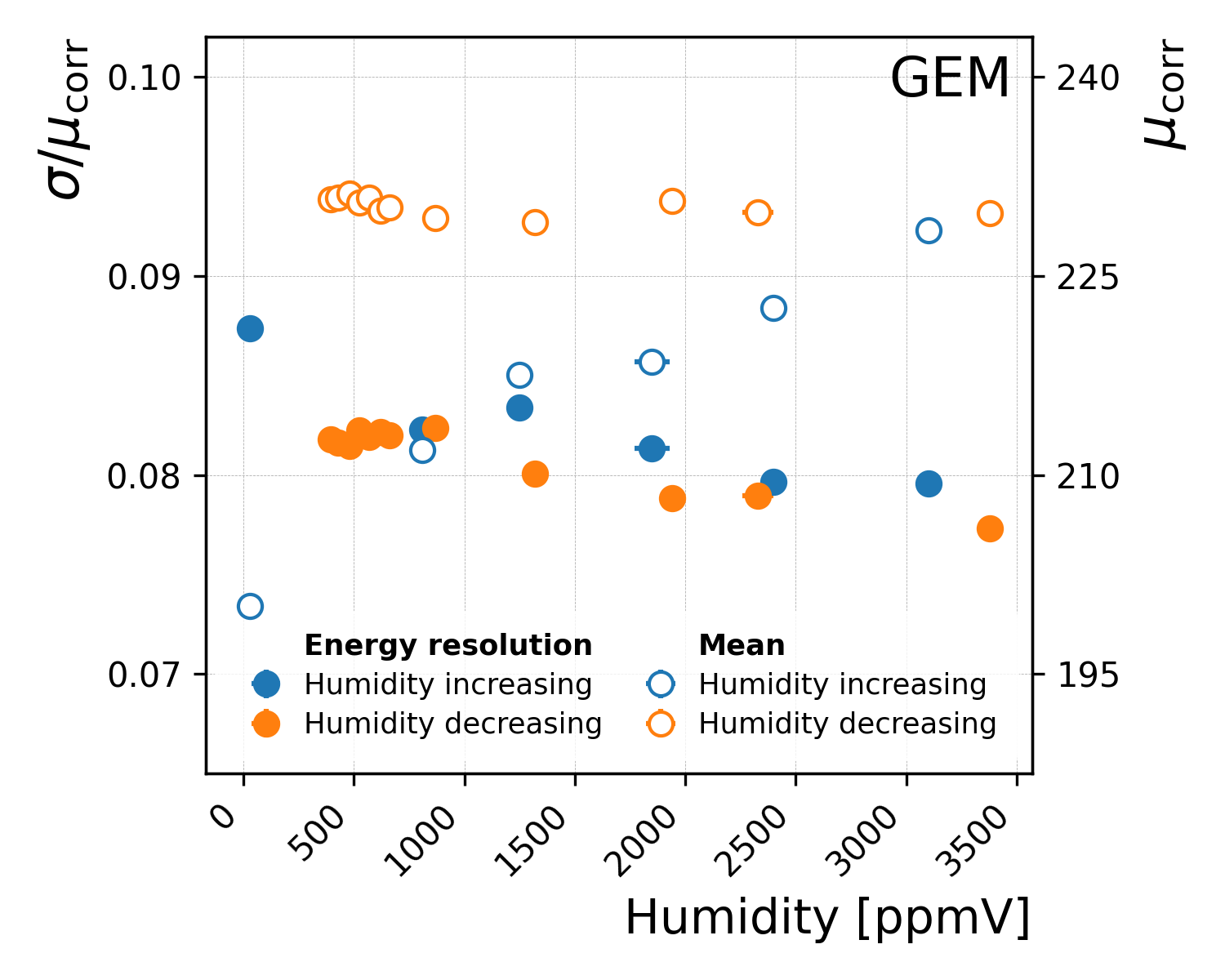}\hfill
    \includegraphics[width=0.49\linewidth]{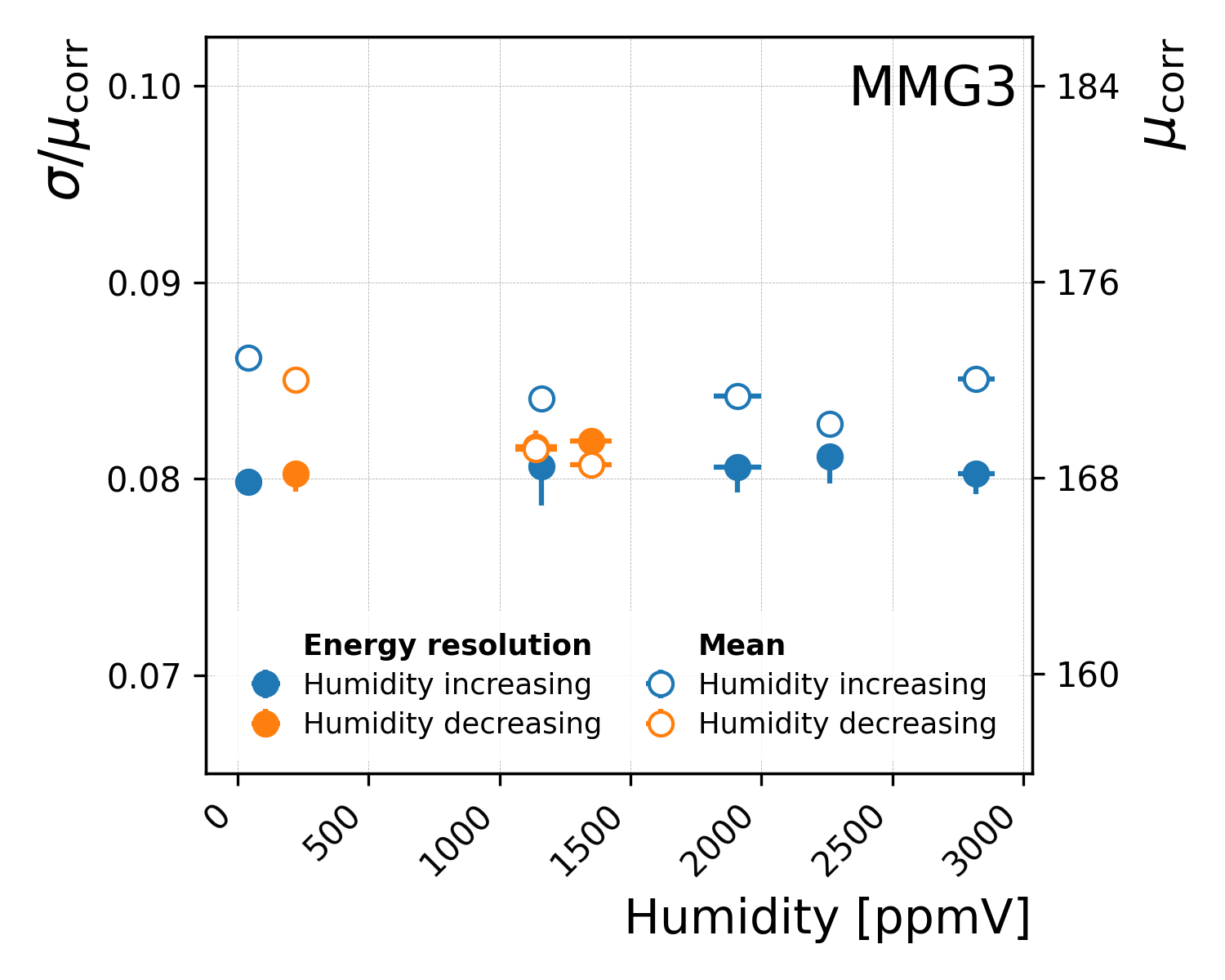}
    \caption{The energy resolution ($\sigma/\mu_{\textrm{corr}}$, left axis, full points) and the main peak position ($\mu_{\textrm{corr}}$, right axis, hollow points) measured with $^{55}$Fe source as a function of absolute humidity for GEM (left) and MMG3 (right) structures. Different colours relate to measurements with increasing or decreasing humidity content in the gas. See text for more details.}
    \label{fig:e_res}
\end{figure*}

The corrected mean peak position ($\mu_{\mathrm{corr}}$) and corresponding energy resolution ($\sigma/\mu_{\mathrm{corr}}$) values are shown in \figref{fig:e_res} for the GEM and Micromegas (type MMG3) on the left and right panels, respectively. In both cases, the results obtained with increasing and decreasing humidity values are shown. 

\begin{figure}[hbt]
    \centering
    \includegraphics[width=\linewidth]{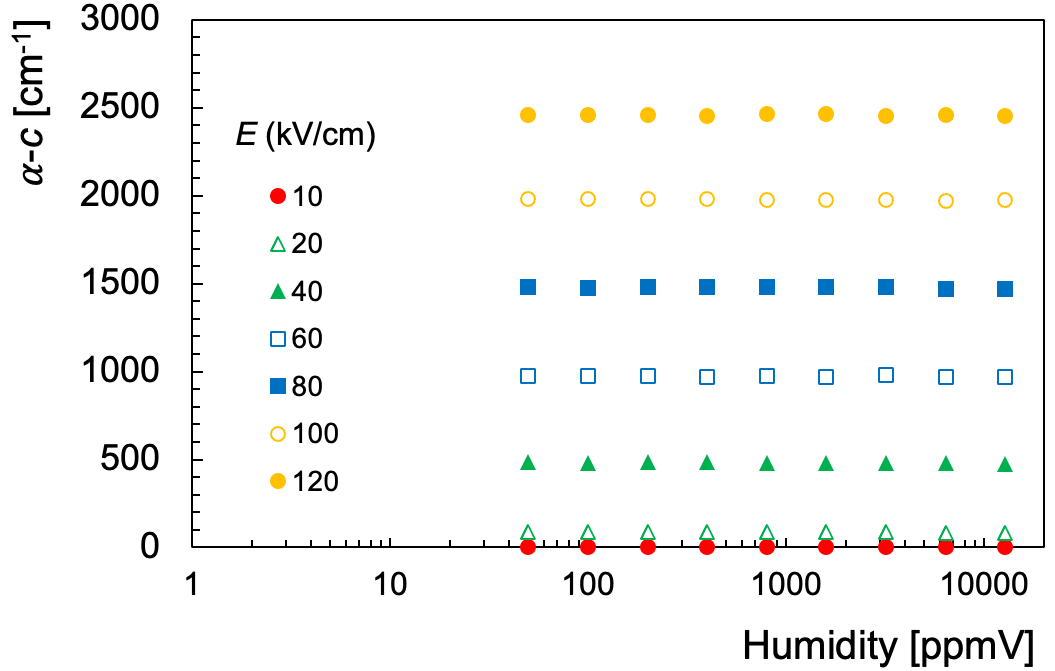}
    \caption{Effective Townsend coefficient of \ArCOtwo mixture calculated with Magboltz~\cite{Biagi1018382} as a function of the humidity content for various electric fields.}
    \label{fig:townsend}
\end{figure}

The GEM gain values increase with humidity in the measured range by up to 10\%, while the MMG3 gain remains relatively stable, or no clear dependency can be observed. 
Indeed, drastic changes in the gas gain with increasing water content are not expected, given the amount of water in the mixture. \Figref{fig:townsend} shows the results of Magboltz~\cite{Biagi1018382} simulations performed for the \ArCOtwo mixture with various absolute humidity levels corresponding to the measured values. The effective Townsend coefficient (i.e. the first Townsend coefficient $\alpha$ minus the attachment $c$) is plotted against the absolute water content for different electric field values, spanning from 10\,kV/cm to 120\,kV/cm (see legend). In all cases, the effective Townsend coefficient values are constant, pointing to marginal (or none) gain dependency in the studied humidity range. The influence of humidity on the Townsend and attachment coefficients may, however, require further investigations. For example, in the studies on the effect of humidity on dielectric breakdown properties of air~\cite{Li_2018} it was shown that the development of electron avalanche should include ion kinetics, where detachments and ion conversions play a significant role in the generation and loss of electrons. In the studies of air with a water content of \SIrange{0}{8}{\percent{}}, it was presented that humidity reduces the effective Townsend coefficient at lower fields, up to $\approx160$\,Td, and increases it at higher fields. The studies presented in ~\cite{Li_2018} show that the full understanding of the amplification and discharge processes in a given gas requires more comprehensive avalanche models considering both spatial growth and temporal processes. Similar considerations may be found attractive for gaseous detector developments and shall be considered in the follow-up studies.

The GEM gain dependency shown in \figref{fig:e_res} is less apparent in the measurement with decreasing humidity. Apart from the residual $T/p$ influence, this may also be related to water absorbed in polyimide during the first measurement series, which influences the electrical and electrostatic properties of the GEM structure (e.g. charging-up behaviour). Indeed, in the case of MMG3, no significant gain dependency and no differences are observed between the measurements with increasing and decreasing humidity content. The Micromegas pillars are made of photopolymer solder mask material (DuPont\textsuperscript{TM} Pyralux\textsuperscript{\textregistered} PC 1000) and cover roughly \SIrange{0.2}{0.5}{\percent{}} of the detector area, contrary to GEMs, where polyimide foil (Apical\textsuperscript{\textregistered}) fills more than 75\% of the active area (see \tabref{tab:mpgd}). Thus, the water uptake by the MMG3 detector shall be largely reduced in comparison to GEMs, which may explain the different results obtained with these two types of detectors. In addition, the hygroscopicity of GEM and MMG3 materials shall be considered, as they can further influence the water trapping in these two structures. It is expected that the water absorption of a polymer mask is lower than that of the Apical\textsuperscript{\textregistered} foil (water absorption of 2.7\%, according to the producer~\cite{Apical200NP}), however, no exact value could be identified for the former. 

Further studies are therefore needed to fully understand the observed dependency and decouple various effects. However, for the sake of the presented studies, the conservative approach is pursued: given a slight influence of gas humidity on gain, the latter is measured for each discharge measurement at a given voltage setting.

For completeness, the effect of gas humidity on energy resolution is studied. It is observed to be very weak for both GEM and MMG3, even though, given the electronegativity of water molecules, one could expect the attachment to deteriorate the energy resolution. As discussed above, the influence of humidity on the effective Townsend coefficient for different gases may vary depending on the electric field strength, however, we do not consider this effect in the current studies. The GEM measurements show a slight improvement in the energy resolution with increasing water content, which may be related to the increasing gain values (see above). Still, the relative change of energy resolution by less than 5\% in the measured humidity range can be considered marginal. 
To summarise, the effect of humidity on the performance of the detector is found to be negligible and should not alter the studies on discharge stability. Of course, a complete interpretation of the observed effects should be considered with a dedicated measurement campaign.

\subsection{Discharge measurement}\vskip5pt

The discharge rate $R_\mathrm{dis}$ is defined as the number of measured discharges $N_\mathrm{dis}$ occurring within the measurement time $t_\mathrm{meas}$. Consequently, the discharge probability $P_\mathrm{dis}$ is given by the ratio of the discharge rate to the measured source rate $R_\mathrm{src}$, following:

\begin{equation}
    P_\mathrm{dis}=\frac{R_\mathrm{dis}}{R_\mathrm{src}}=\frac{N_\mathrm{dis}}{R_\mathrm{src}\cdot t_\mathrm{meas}}
\end{equation}

\begin{figure*}[hbt]
    \centering
    \includegraphics[width=.4\textwidth]{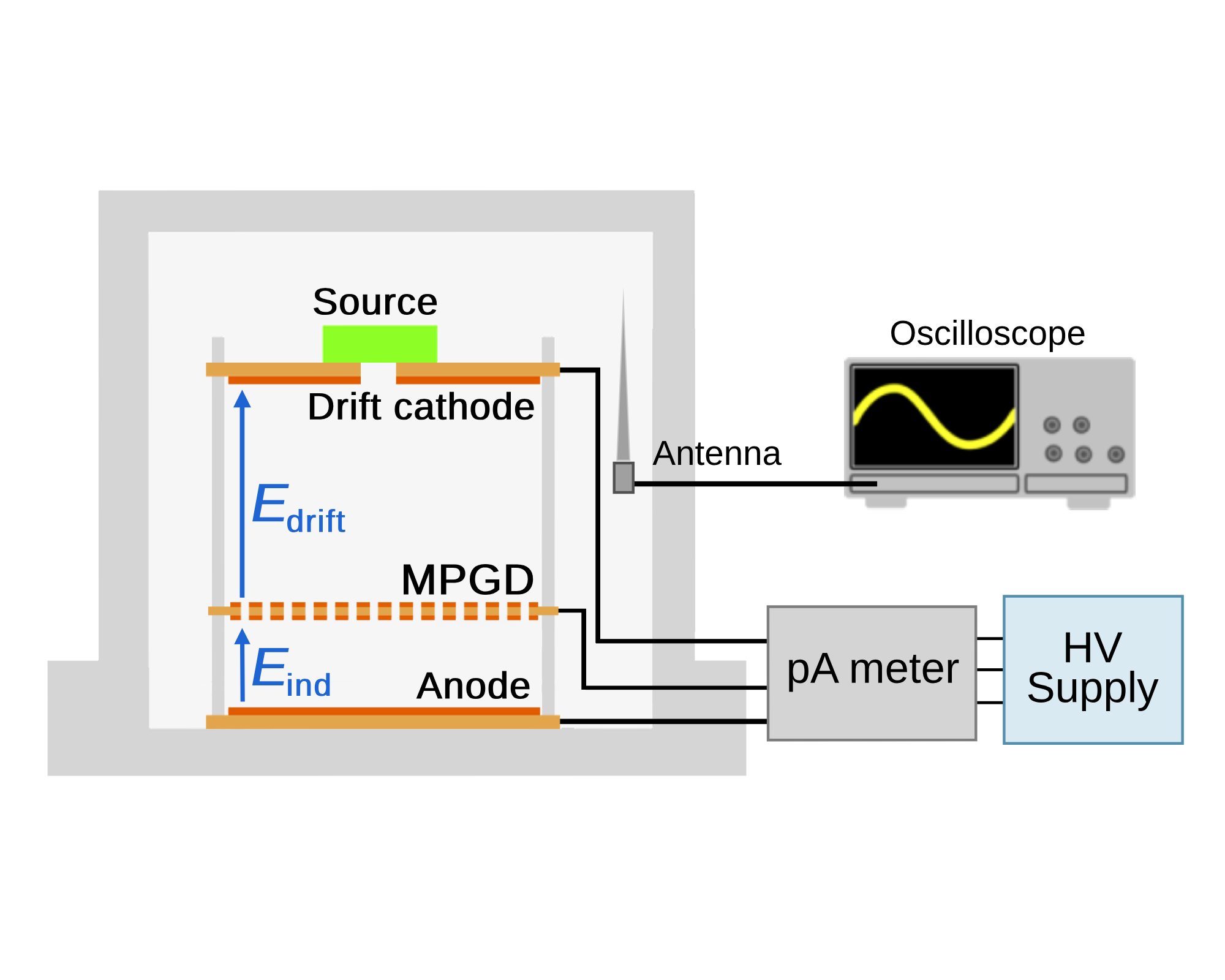}
    \includegraphics[width=.4\textwidth]{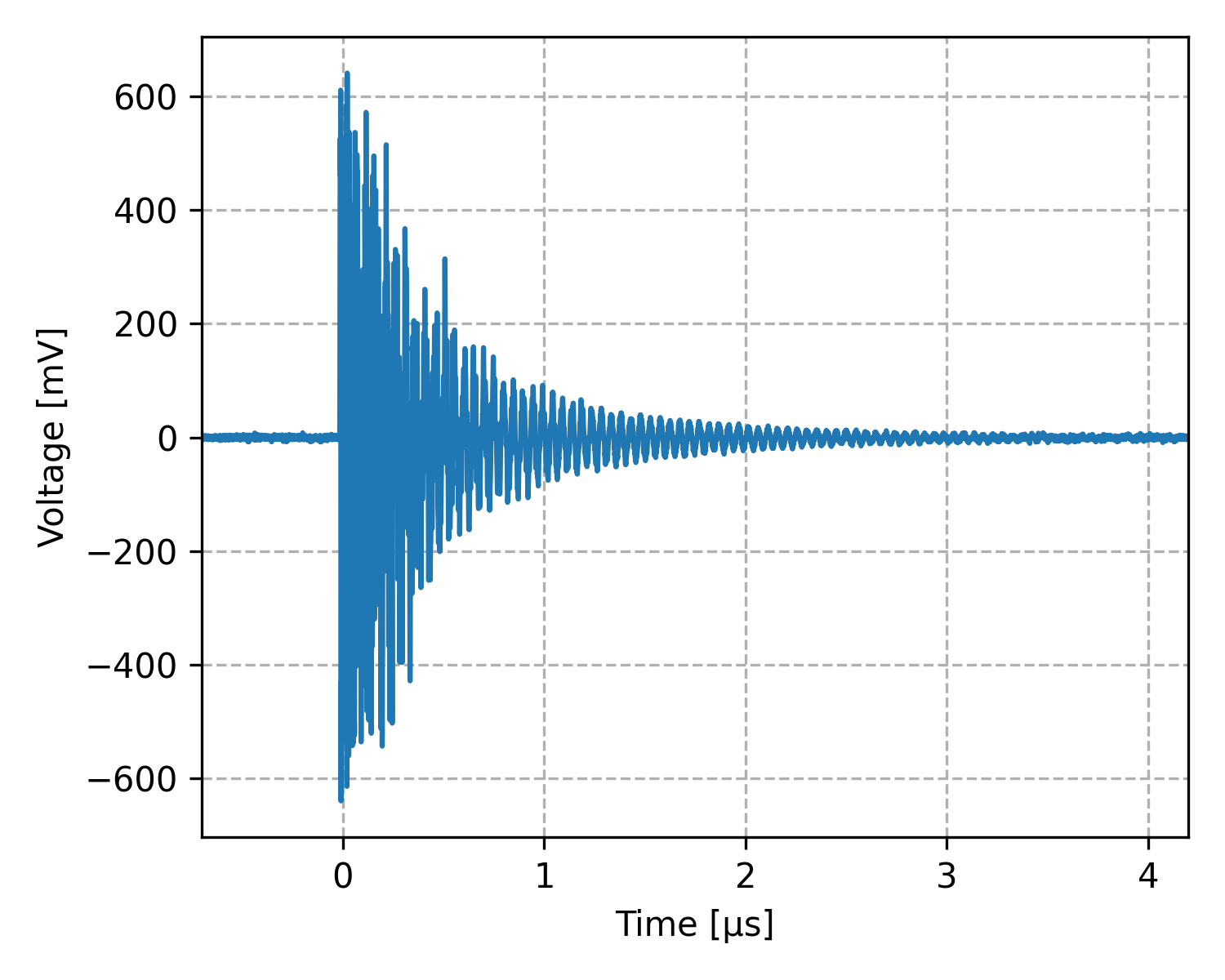}
    \caption{(left) Schematic illustration of the detector’s electrical connections. (right) A typical discharge signal obtained from a THGEM and recorded by an oscilloscope via the antenna installed in the gas vessel.}
    \label{fig:discharge_signal}
\end{figure*}

Discharges are counted by placing an open-ended cable within the detector chamber, where it functions as a simple antenna. The cable is connected to an oscilloscope, which records and displays the discharge signals, as shown in \figref{fig:discharge_signal}.

To determine the discharge probability as a function of gain, both quantities must be measured simultaneously. The pico-amperemeter, used to measure the amplification current (see \secref{sec:gain}), registers also discharge events as sharp spikes, altering the gain estimation. To prevent this, an averaging algorithm excludes points that deviate more than $5\sigma$ from the mean current value. The procedure was used in our previous studies and is explained in more detail in ~\cite{Gasik:2022vft}. In each measurement of the discharge probability, up to hundreds of sparks are recorded with the oscilloscope. However, the discharge rate never exceeds \SI{1}{\Hz} to assure reliable gain measurement. The associated statistical error of discharge occurrences is calculated by assuming a Poisson distribution of the latter.

\section{Results}
\subsection{Gain comparisons}\vskip5pt
\label{sec:res:gain}

\begin{figure*}[ht]
    \centering
    \includegraphics[width=0.8\linewidth]{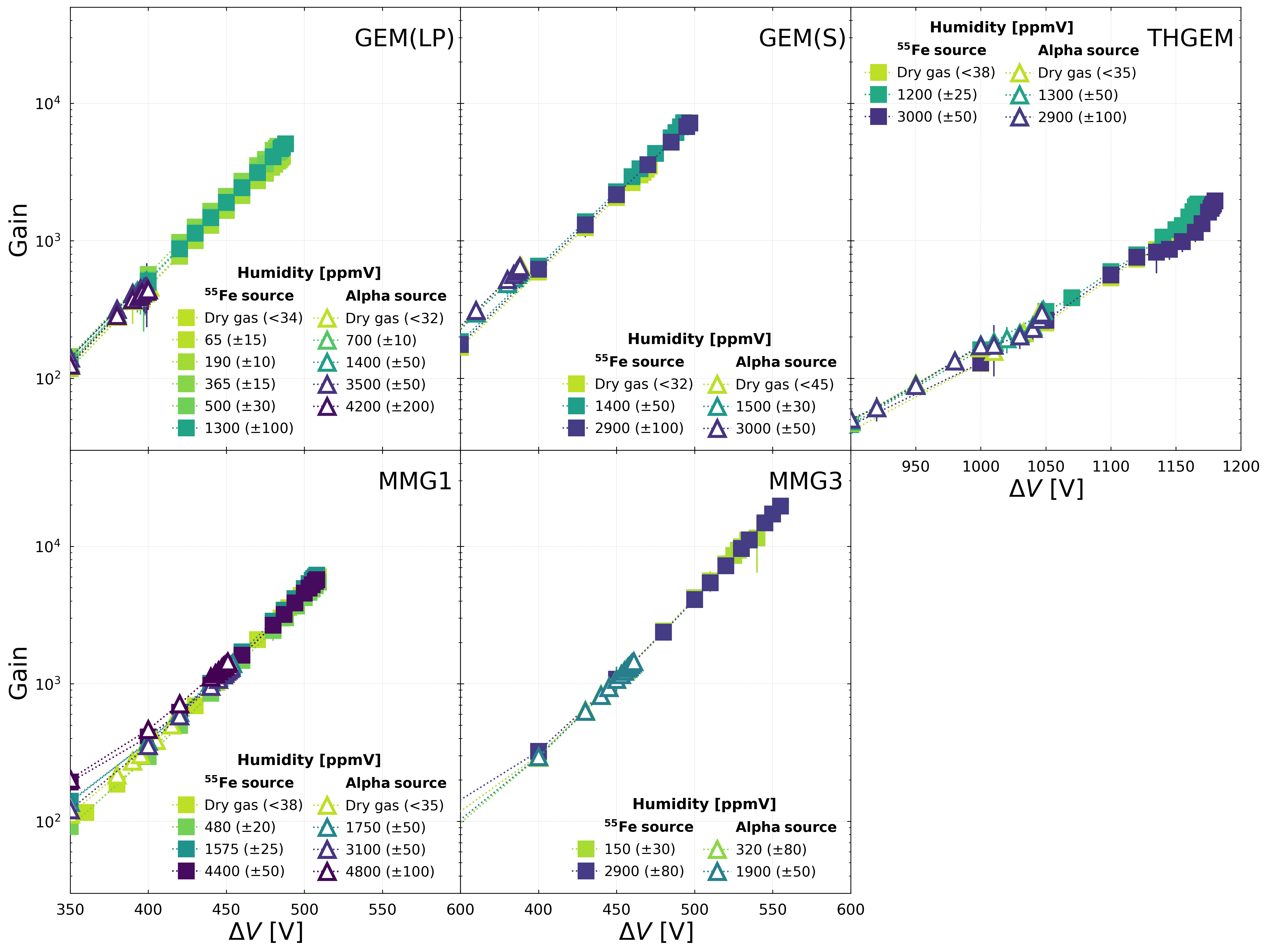}
    \caption{The absolute gain as a function of the applied voltage measured with various MPGDs, in \ArCOtwo mixture. The different colours display the humidity content of the gas mixture according to the legend. The gain values of MMG1 are corrected for the $<$100~\% collection efficiency --- see \secref{sec:gain}.}
    \label{fig:gain}
\end{figure*}

\begin{figure*}[h!]
    \centering
    \includegraphics[width=0.8\linewidth]{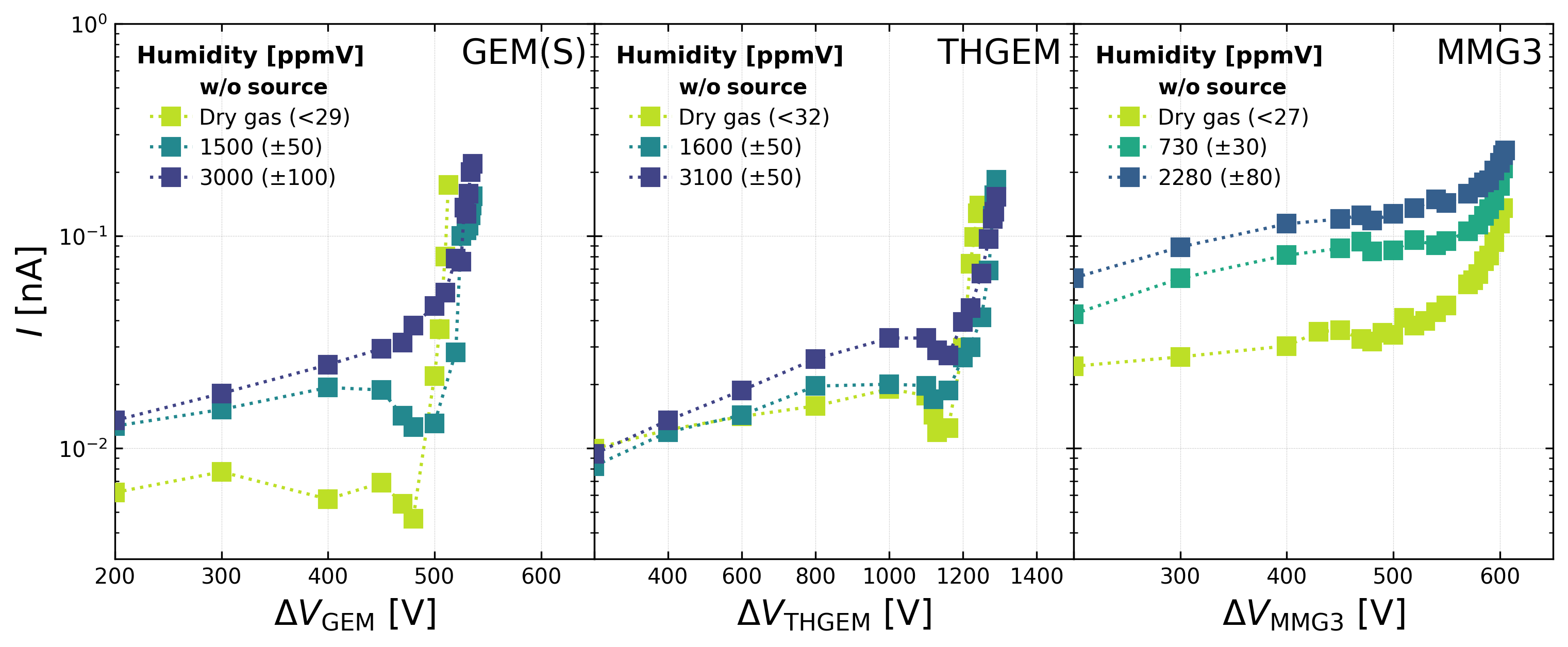}
    \caption{Leakage current of a standard GEM (left), a Thick GEM (middle) and a Micromegas (right) detector measured as a function of the amplification voltage for various levels of humidity in \ArCOtwo.}
    \label{fig:ileak}
\end{figure*}

\Figref{fig:gain} shows the measured absolute gain as a function of the amplification voltage applied to each MPGD. The two different sources used for these studies, the alpha source and the $^{55}$Fe source (see \secref{sec:src}), are presented as triangles and squares, respectively. The humidity content of the gas mixture is indicated by the changing colour. 
To minimise the influence of humidity variations on measurements presented in this and the following sections, the humidity is adjusted and stabilised overnight. Thus, the humidity level in the gas mixture can be considered constant for a given gain or discharge probability curve.

As expected, the gain dependence on the applied voltage follows an exponential behaviour. A sudden drop of the THGEM gain at higher voltages is a known effect linked to the onset of discharges and discussed in more detail in~\cite{Gasik:2022vft}. When comparing the measurements between different MPGDs and radioactive sources, it can be clearly seen that varying humidity levels in the gas do not significantly affect the absolute gain. Also, no clear dependency can be observed, contrary to the discussion in \secref{sec:adc}. However, the measured gain values are not corrected for $T/p$ variations, which, given that each measurement is performed on a different day, can dissolve the already vague dependency. 

\subsection{Leakage current studies}\vskip5pt
The influence of humidity on the leakage current of an MPGD structure is presented in \figref{fig:ileak}. The measurements are performed without any radioactive source. All data is collected after the gas mixture's humidity level stabilises. For the measurements with dry gas, the time elapsed since the beginning of flushing is long enough (in all cases, this happened overnight) to assume that most of the humidity trapped in the material of the detector is removed. 

For the three MPGD structures studied, the leakage current is clearly dependent on gas humidity. At very high voltages, specifically those exceeding the operational range used for gain measurements (where the discharge rate with radioactive sources becomes excessively high), the current-to-voltage relationship deviates from linearity. In this regime, the influence of gas humidity on leakage current is significantly reduced. The latter is even reversed in the case of GEM and THGEM, as the highest leakage current values are measured for the dry gas. This can be related to the onset of discharges that are likely to appear at very high fields, especially near the electrode defects. This will be further investigated in the discharge stability studies, discussed in the following \secref{sec:res:disch}.\\

The study shows that humidity can alter the leakage current measurement, however, in the measured range, the differences are marginal. In all cases, the leakage current values measured below the high-voltage/high-current region are at the level of tens of picoamperes. This is well below the limits used in the quality control of MPGDs produced for large-scale projects like ALICE TPC~\cite{Adolfsson_2021}, CMS GE1/1 Muon~\cite{ABBAS2022166716} and ATLAS New Small Wheel~\cite{ALLARD2022166143} upgrades during the LHC Long Shutdown 2. Indeed, in all three projects, the HV stability and leakage currents of MPGD structures were measured, requiring relative humidity of the gas below 10\%, without specifying the exact value. This is well in line with the range considered in this work. As mentioned in \secref{sec:intro}, various studies showed that at RH values above 50\% clear instabilities can be observed in the form of enhanced discharge rate or leakage (dark) current values. This region of humidity content is not to be discussed here.

\subsection{Discharge stability studies}\vskip5pt
\label{sec:res:disch}
\begin{figure*}[h]
    \centering
    \includegraphics[width=0.8\linewidth]{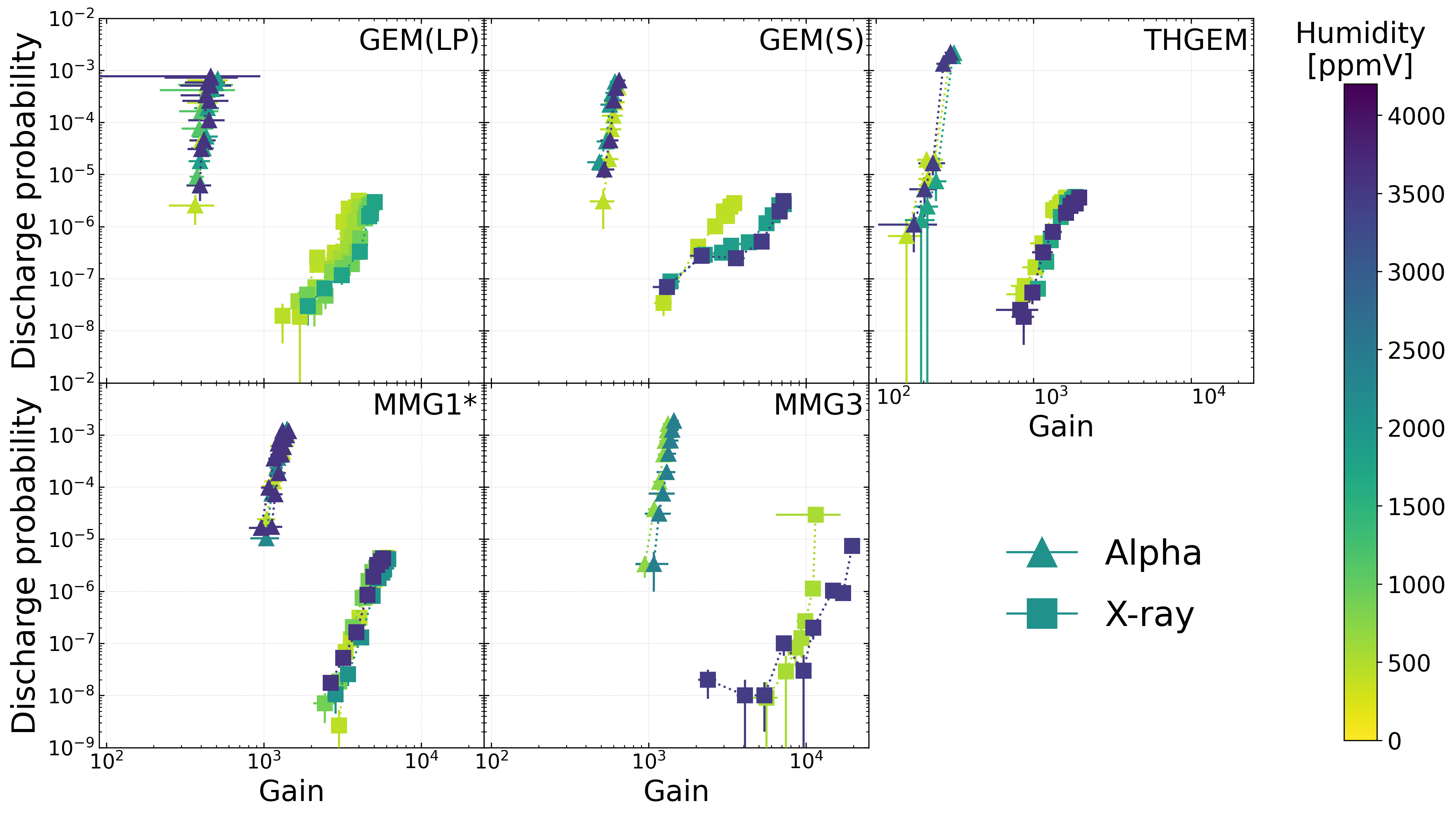}
    \caption{The discharge probability as a function of the measured gain for each MPGD and the different sources --- see legend. (*) The gain values of MMG1 are corrected for the $<$100~\% collection efficiency --- see \secref{sec:gain}.}
    \label{fig:disprob}
\end{figure*}

\begin{figure*}[h!]
    \centering
    \includegraphics[width=0.8\linewidth]{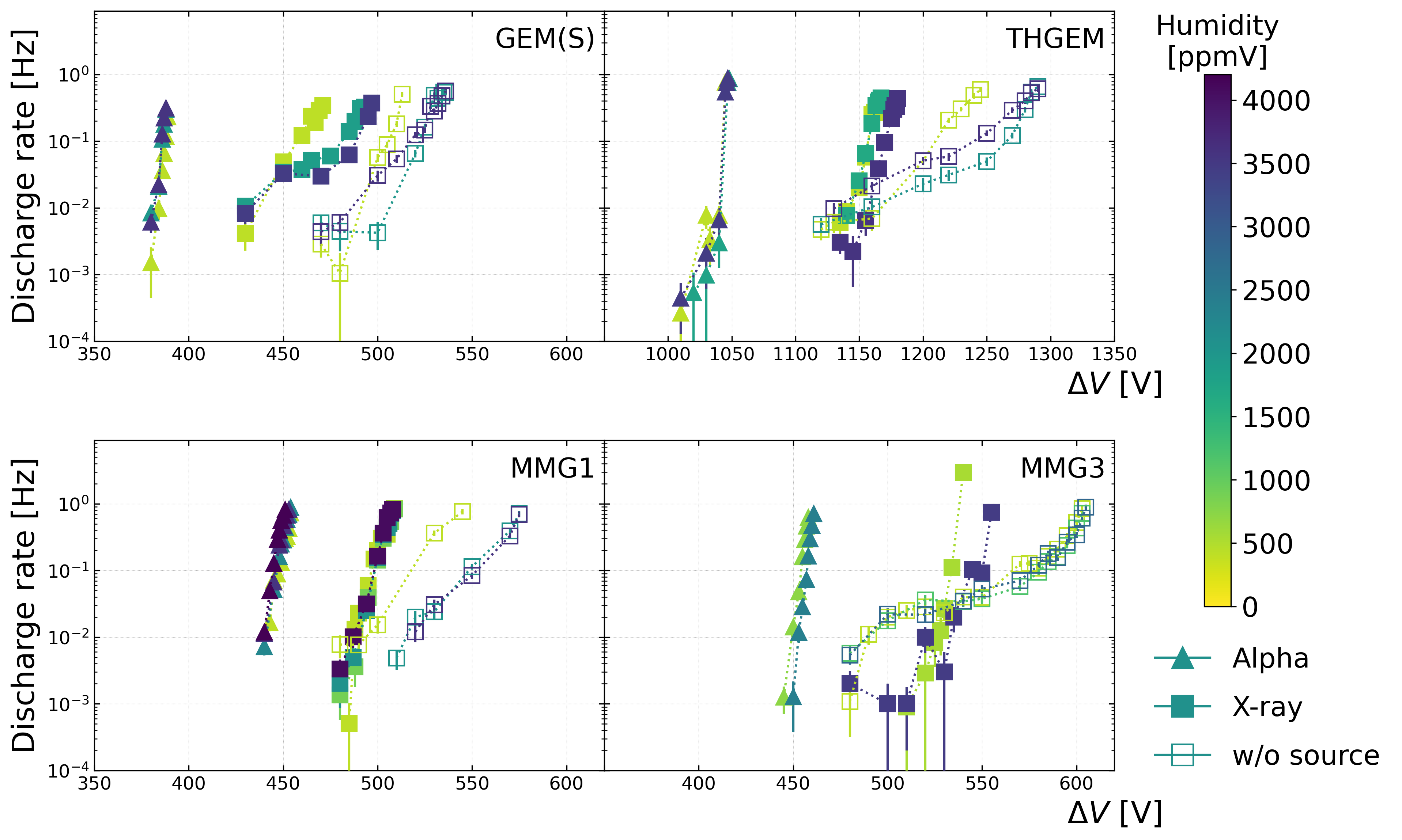}
    \caption{The discharge rate as a function of the amplification voltage for each MPGD measured with and without a source. All measurements are performed in \ArCOtwo gas mixture. The humidity content of the gas mixture follows the colour bar.}
    \label{fig:disrate}
\end{figure*}

\Figref{fig:disprob} shows the discharge probability of the MPGDs using the different sources as a function of the absolute gain. The colour bar indicates the different humidity levels in the gas mixture, while different symbols represent different radioactive sources. The large vertical error bars (discharge probability) arise from the limited number of discharge events --- typically fewer than ten --- recorded during a single 900~s measurement, constrained by the requirement to maintain stable conditions. Regarding the horizontal error bars, in the high-gain regime, the frequent discharging (above 0.3 Hz) makes it difficult to find sufficiently long, stable intervals for reliable current measurements. The limited number of measurement points within these time windows leads to large statistical uncertainties in the gain determination, reflected in the size of error bars.

While comparing the results obtained with alpha particles, one can immediately observe that the discharge probability scales inversely with the (TH)GEM hole or MMG cell pitch, thus the hole/cell density. As we have shown in~\cite{GASIK2017116, Gasik:2022vft, Gasik_2023} this effect can be explained by simple consideration of a primary charge density arriving at the single amplification cell. Assuming 100\% electron collection efficiency (in the case of the MMG1, the gain is corrected for the $<$100\% value --- see \secref{sec:gain}), the number of primary electrons entering a single amplification hole/cell increases with decreasing hole/cell density. Thus, given the gain value, the probability of reaching the critical charge required for a streamer and spark development is increasing~\cite{GASIK2017116, Gasik:2022vft, Gasik_2023}. 
Measurements with alphas are particularly sensitive to these effects as primary charge densities are very large, and the critical charge limits can be reached already at relatively low gains, where other effects related to, for example, defects, high electric fields at the electrode edges, or contamination, are not yet noticeable. 

The density effect is evident in both Micromegas detectors when compared to the (TH)GEM structures. However, interpreting the results obtained with both meshes is more complex. A clear discharge probability dependency on the mesh cell size (larger mesh density results in lower discharge rate) was observed before in ~\cite{ALVIGGI2020162359} and has been shown again and discussed in~\cite{Gasik_2023}, where the conclusion of considering Micromegas cells as independent amplification unit (as a GEM hole) was proposed. Comparing present results obtained with alpha particles for MMG1 and MMG3 is not straightforward, though. As all the measurements presented in this work are performed with a drift field of 400\,V/cm, the collection efficiency for the MMG1 is lower than 80\%~\cite{Gasik_2023} unlike for all other structures, where the value of 100\% can be safely assumed (see \secref{sec:gain}). The gain values of MMG1 in \figref{fig:gain} and \figref{fig:disprob} have been corrected for this inefficiency. In reality, less charge enters the amplification gap in MMG1 compared to MMG3, which in turn affects the discharge probability, which does not scale linearly with gain. Therefore, great care must be taken when comparing discharge probability curves between MMG1 and MMG3 in \figref{fig:disprob}.

Discharge probability measurements with the $^{55}$Fe source require higher gains as the primary charge densities are much lower than those provided with an alpha source. Thus, the multiplication required to reach the critical charge limits is proportionally higher. The GEM hole or MMG cell density effect can still be recognised in the measurements with $^{55}$Fe. However, it is less pronounced when comparing the structures with similar geometries (pitches), such as the standard and the LP GEMs. We assume that this occurs because additional factors may influence instabilities at higher gains (and voltages), such as the effects of contamination, defects, or electric field hot spots, particularly around electrode edges. As explained above, the MMG1 efficiency correction may not be sufficient to accurately scale the discharge curve, hence, the direct comparison of the MMG1 and MMG3 curves does not provide a strong basis for drawing definitive conclusions.

When comparing the discharge probabilities obtained with the alpha source measurements for all of the MPGDs, no hierarchy is observed in terms of different humidity contents. In the case of the THGEM and the Micromegas, the same observation is made when measuring with the $^{55}$Fe source. However, the results obtained with the standard and large-pitch GEMs and the $^{55}$Fe-source show a variation in which, at higher gains, the discharge probability decreases when adding humidity to the gas. A similar conclusion can be drawn from the results of MMG3, however, here the difference between the two measurements (dry and humidified gas) concerns a single point, thus making the observation less significant.

It can be concluded that, as stated before, the humidity effects do not matter when we discuss the intrinsic properties of the gas mixture (critical charge limit resulting in a streamer development), which are manifested best in studies with alpha particles (see above). This is no surprise given the marginal change of the gas mixture itself, as it was also discussed in the amplitude studies in \secref{sec:adc}.
However, with effects which may appear at higher voltages, the water content seems to play a noticeable role. In order to study this effect further, a measurement of the detector stability at very high voltages is performed. In order to reach the very high voltage region, no source is used to induce the discharges.

\Figref{fig:disrate} illustrates the discharge rate as a function of the applied amplification voltage for various MPGDs and humidity levels. In addition to the data presented in \figref{fig:disprob}, this plot also includes measurements taken without any radiation source (hollow points). In this configuration, the observed discharges result from cosmic rays, field emissions, and other factors rather than source irradiation and are referred to as spurious discharges. The data points corresponding to measurements without a source are represented by hollow squares and extend to higher voltages due to reduced primary ionisation.
Notably, these measurements reveal a clear correlation between increasing humidity content in the gas mixture and discharge behaviour. Increased stability against spurious discharges is observed in the GEM and THGEM, particularly at the highest voltages. A similar trend is seen in Micromegas, where this dependence is also evident at lower voltages. Additionally, the measurements appear to group at the highest voltages when humidity is introduced, irrespective of its absolute value. This suggests that beyond a certain point, adding more water vapour does not further enhance MPGD stability. These findings indicate that humidity has minimal impact on discharge formation at lower voltages but plays a role in reducing spurious discharges, potentially by mitigating electrode defects or influencing the charging-up of insulating layers.

One should consider the fact that measurements without a source may be biased by missing gain estimates. From the previous measurements (see \figref{fig:gain}), however, we observe the gain should not change by more than \SIrange{10}{20}{\percent{}}. 
Varying ambient conditions shall be taken as an additional uncertainty. Still, in all observed cases, the dry mixtures seem to be less stable in terms of spurious discharge probability, which allows us to conclude that this is a genuine, systematic effect.

\section{Conclusion}
The impact of humidity on the performance of three different non-resistive MPGD structures (GEM, Thick GEM, and Micromegas) is studied using humidified \ArCOtwo gas mixture from a dedicated gas system. To evaluate the performance of the detector, different properties are investigated, including gain, energy resolution, leakage current, discharge rate and discharge probability. The measurements are performed both with (alpha, \xray) and without radiation sources.

Studies on gain and energy resolution indicate that humidity has a negligible impact on detector performance and should not affect investigations on discharge stability. Indeed, the effective Townsend coefficient calculated with Magboltz shows no dependence on the water content within the studied humidity and electric field ranges. However, a complete understanding of this behaviour may require more sophisticated avalanche modelling, including, for example, ion kinetics. The residual gain variations observed in the $^{55}$Fe peak position studies and the absolute gain measurements may also be related to the ambient conditions or water absorption in the polyimide material of GEMs. Full interpretation of the effects observed in this work should be considered with a dedicated measurement campaign. 

The leakage current shows a dependence on humidity, which may be attributed to water conductivity and moisture accumulation on insulating surfaces such as the pillars of the Micromegas, the polyimide in GEMs, or the glass-reinforced epoxy laminate used in the Thick GEM structure. However, within the operational voltage range, these variations remain minimal. At very high amplification fields, the leakage current increases, though without a clear correlation to humidity. This rise may be associated with the occurrence of spurious discharges, which are more likely to develop at extremely high fields, particularly near electrode defects.

Finally, we observe no major influence of water content on the discharge stability of any of the amplification structures when discharge probability is measured with an alpha or an \xray source in \ArCOtwo mixture. However, the measurements with $^{55}$Fe at the highest amplification fields show a slight dependency towards the higher stability of a humid gas. This observation is further confirmed when the discharge rate is measured without any radiation source. In all the studied structures, we observe enhanced stability (lower discharge rates) when the detector is operated with humidified gas. However, no major improvement is observed when the water content is further increased. These findings suggest that the addition of a small amount of water vapour in the range of \SIrange[range-units=single]{100}{1000}{ppm_V} can effectively enhance discharge stability. 
Nevertheless, the optimal humidity level should be determined individually for each application and gas mixture, taking into account whether the aim is to suppress leakage currents, mitigate discharge instabilities, or address other specific operational requirements of the detector.

In conclusion, while humidity does not appear to influence the intrinsic discharge formation mechanism, it does reduce the rate of spurious discharges, likely associated with electrode imperfections or the charging-up of insulating surfaces.

\section*{Acknowledgements}
The studies have been performed in the framework of the RD51 and DRD1 Collaborations. 

We warmly thank C.~Garabatos for conceptualising the gas system used to humidify the gas mixture, providing Magboltz calculations, and for many fruitful discussions. We also warmly thank R. de Oliveira for providing technical information on the used MPGDs.

This work was supported by the Deutsche Forschungsgemeinschaft - Sachbeihilfe [grant number DFG FA 898/5-1].

\balance

\bibliography{./references.bib}

\end{document}